\documentclass[twocolumn,twoside,slac_two]{revtex4}
\usepackage{graphicx}
\usepackage{fancyhdr}
\begin{document}

\title{Probing the Quark-Gluon Phase Transition with Correlations and Fluctuations in Heavy Ion Collisions from the STAR Experiment}

%

\author{Terence J. Tarnowsky (for the STAR Collaboration)}
\affiliation{National Superconducting Cyclotron Laboratory, Michigan State University, East Lansing, MI 48824, USA}

\begin{abstract}
The measurement of particle correlations and fluctuations has been suggested as a method to search for the existence of a phase transition in relativistic heavy ion collisions. If quark-gluon matter is formed in the collision of relativistic heavy ions, measuring these correlations could lead to a determination of the presence of partonic degrees of freedom within the collision. Additionally, non-statistical fluctuations in global quantities such as baryon number, strangeness, or charge may be observed near a QCD critical point.
Results for short and long-range multiplicity correlations (forward-backward) are presented for several systems (Au+Au and Cu+Cu) and energies (e.g. $\sqrt{s_{NN}}$ = 200, 62.4, and 22.4 GeV). For the highest energy central A+A collisions, the correlation strength maintains a constant value across the measurement region. In peripheral collisions, at lower energies, and in pp data, the maximum appears at midrapidity. Comparison to models with short-range (HIJING) and both short and long-range interactions (Parton String Model) do not fully reproduce central Au+Au data. Preliminary results for K/$\pi$ fluctuations are also shown as a function of centrality in Cu+Cu collisions at $\sqrt{s_{NN}}$ = 22.4 GeV. 

\end{abstract}

\maketitle

\thispagestyle{fancy}


\section{Introduction}

Fluctuations and correlations are well known signatures of phase transitions. In particular, the quark/gluon to hadronic phase transition may lead to significant fluctuations. As part of a proposed beam energy scan at the Relativistic Heavy Ion Collider (RHIC), the search for the QCD critical point will make use of the study of correlations and fluctuations, particularly those that could be enhanced during a phase transition that passes close to the critical point. Forward-backward (FB) multiplicity correlations and particle ratio fluctuations are two observables that have been studied as a function of energy and system-size. They will also provide additional information to assist in locating the QCD critical point when measured as part of the proposed beam energy scan.

Forward-backward multiplicity correlations have been studied in elementary particle collisions \cite{1,2}, and more recently, from the STAR experiment in heavy ion collisions at RHIC \cite{FB, CPOD, bks_QM2008, tt_thesis, fb_prl}. FB correlations are measured as a function of pseudorapidity ($\eta$). Because they probe the longitudinal characteristics of the system produced in heavy ion collisions, they provide unique insight into the space-time dynamics and the earliest stages of particle production. A relationship between the multiplicity in the forward and backward hemisphere was found in hadron-hadron collisions \cite{fblinear},

\begin{equation}\label{linear} 
<N_{b}(N_{f})> = a + bN_{f} 
\end{equation}

where the coefficient {\it b} is referred to as the correlation strength. Eq. \ref{linear} can be expressed in terms of the expectation values,

\begin{equation}\label{dispersion}
b = \frac{<N_{f}N_{b}>-<N_{f}><N_{b}>}{<N_{f}^{2}>-<N_{f}>^{2}} = \frac{D_{bf}^{2}}{D_{ff}^{2}}
\end{equation}

where $D_{bf}^{2}$ and $D_{ff}^{2}$ are the backward-forward and forward-forward dispersions, respectively. All quantities in Eq.~(\ref{dispersion}) are measurable experimentally. The presence of long-range FB correlations has been predicted in some Monte Carlo particle production models, specifically the Dual Parton Model (DPM) and the Color Glass Condensate/Glasma (CGC) model \cite{DPM, CGC}. Both models contain a form of longitudinal color flux tubes: color strings in the DPM and Glasma flux tubes in the CGC \cite{DPM_CGC}. These longitudinal structures are considered the origin of the long-range correlation.

Dynamic particle ratio fluctuations, specifically fluctuations in the $K/\pi$ ratio, can provide information on the quark-gluon to hadronic phase transition. These can be measured using the variable $\nu_{dyn}$, originally introduced to study net charge fluctuations \cite{nudyn1, nudyn2}. $\nu_{dyn}$ quantifies deviations in the particle ratios from those expected for an ideal statistical Poissonian distribution. The definition of $\nu_{dyn,K/\pi}$ (describing fluctuations in the $K/\pi$ ratio) is,

\begin{eqnarray}
\nu_{dyn,K/\pi} = \frac{<N_{K}(N_{K}-1)>}{<N_{K}>^{2}}      \nonumber \\
+ \frac{<N_{\pi}(N_{\pi}-1)>}{<N_{\pi}>^{2}}
- 2\frac{<N_{K}N_{\pi}>}{<N_{K}><N_{\pi}>}
\label{nudyn}
\end{eqnarray}

where $N_{K}$ and $N_{\pi}$ are the number of kaons and pions in a particular event, respectively. In this proceeding, $N_{K} = N_{K^{+}+K^{-}}$ and $N_{\pi} = N_{\pi^{+}+\pi^{-}}$ are the total charged multiplicity for each particle species. $\nu_{dyn}$ = 0 for the case of a Poisson distribution of kaons and pions and is largely independent of detector acceptance and efficiency in the region of phase space being considered \cite{nudyn2}. An in-depth study of $K/\pi$ fluctuations in Au+Au collisions at $\sqrt{s_{NN}}$ = 200 and 62.4 GeV was carried out by the STAR experiment \cite{starkpiprl}.

\section{Analysis}

The data presented in this proceeding was acquired by the STAR experiment at RHIC from minimum bias (MB) Au+Au and Cu+Cu collisions \cite{STAR}. The results for FB correlations are from: Au+Au collisions at a center-of-mass collision energy ($\sqrt{s_{NN}}$) of 200 and Cu+Cu collisions at $\sqrt{s_{NN}}$ = 200, 62.4, and 22.4 GeV. The main particle tracking detector at STAR is the Time Projection Chamber (TPC) \cite{STARTPC}. All charged particles in the TPC pseudorapidity range $|\eta| < 1.0$ and with $p_{T} > 0.15$ GeV/c were considered. The $\eta$ range was subdivided into forward and backward measurement intervals of width 0.2$\eta$. Forward-backward correlations were measured symmetrically about $\eta = 0$ with increasing $\eta$ gaps (measured from the center of each bin), $\Delta\eta =$ 0.2, 0.4, 0.6, 0.8, 1.0, 1.2, 1.4, 1.6, and 1.8 units in pseudorapidity. This is different than the standard two-particle correlation function, which uses a relative $\eta$ difference between particle pairs. The minimum bias collision centralities were determined by an offline cut on the TPC charged particle multiplicity within the range $|\eta| < 0.5$. The centralities used in this analysis account for 0-10, 10-20, 20-30, 30-40, and 40-50\% of the total hadronic cross section. A cut on the longitudinal position of the collision vertex ($v_{z}$) to within $\pm 30$ cm from $z=0$ (center of the TPC) was used. Corrections for detector acceptance and tracking efficiency were carried out using a Monte Carlo event generator and propagating the simulated particles through a GEANT representation of the STAR detector geometry. 

\begin{figure}
\centering
\includegraphics[width=90mm]{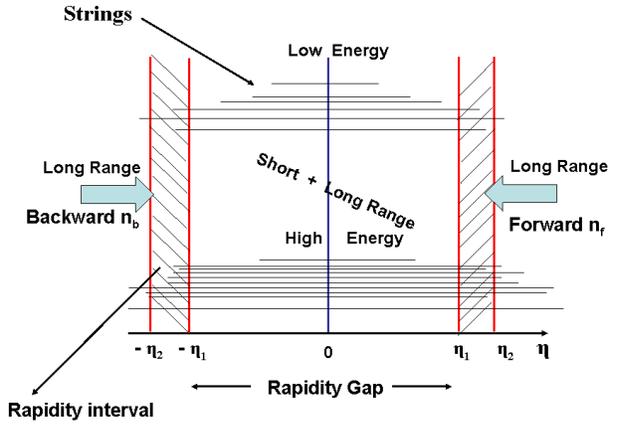}
\caption{Cartoon diagram of a forward-backward correlation measurement.}
\label{fig1}
\end{figure}

To reduce the effect of impact parameter (centrality) fluctuations on this measurement, each relevant quantity ($\left<N_{f}\right>$, $\left<N_{b}\right>$, $\left<N_{f}\right>^{2}$, and $\left<N_{f}N_{b}\right>$) was obtained on an event-by-event basis as a function of STAR reference multiplicity, $N_{ch}$. A linear fit to $\left<N_{f}\right>$ and $\left<N_{b}\right>$, or a second order polynomial fit to $\left<N_{f}\right>^{2}$ and $\left<N_{f}N_{b}\right>$ was used to extract these quantities as functions of $N_{ch}$. Tracking efficiency and acceptance corrections were applied to each event. One correction factor, calculated for each centrality, is applied to the values $\left<N_{f}\right>, \left<N_{b}\right>, \left<N_{f}\right>^{2}$, and $\left<N_{f}N_{b}\right>$ for every event in that centrality bin. Therefore, all events within a particular centrality bin have the same correction factor. These were then used to calculate the backward-forward and forward-forward dispersions, $D_{bf}^{2}$ and $D_{ff}^{2}$, binned according to the STAR centrality definitions and normalized by the total number of events in each bin. A strong bias exists if the centrality definition overlaps with the measurement interval in $\eta$ space. Therefore, there are three $\eta$ regions used to determine reference multiplicity: $|\eta| < 0.5$ (for measurements of $\Delta\eta > 1.0$), $0.5 < |\eta| < 1.0$ (for measurements of $\Delta\eta < 1.0$), or $|\eta| < 0.3 + 0.6 < |\eta| < 0.8$ (for measurement of $\Delta\eta = 1.0$).

For the study of $K/\pi$ fluctuations, the results discussed here are from minimum bias Cu+Cu collisions at $\sqrt{s_{NN}}$ = 22.4 GeV. A cut on the longitudinal position of the collision vertex ($v_{z}$) restricted events to within $\pm 30$ cm of the STAR TPC. All charged particles in the pseudorapidity interval $|\eta| < 1.0$ were measured. The transverse momentum ($p_{T}$) range for pions and kaons was $0.2 < p_{T} < 0.6$ GeV/c. Charged particle identification involved measured ionization energy loss (dE/dx) in the TPC gas and total momentum ($p$) of the track. The energy loss of the identified particle was required to be less than two standard deviations (2$\sigma$) from the predicted energy loss of that particle at a particular momentum. A second requirement was that the measured energy loss of a pion/kaon was more than 2$\sigma$ from the energy loss prediction of a kaon/pion. A third requirement was used to veto electron candidates. Identified pions or kaons with measured energy losses within 1$\sigma$ of predicted energy loss of an electron were rejected. The centralities used in this analysis account for 0-10, 10-20, 20-30, 30-40, 40-50, and 50-60\% of the total hadronic cross section.

\section{Results}
\subsection{Forward-Backward Multiplicity Correlations}

\begin{figure}
\centering
\includegraphics[width=80mm]{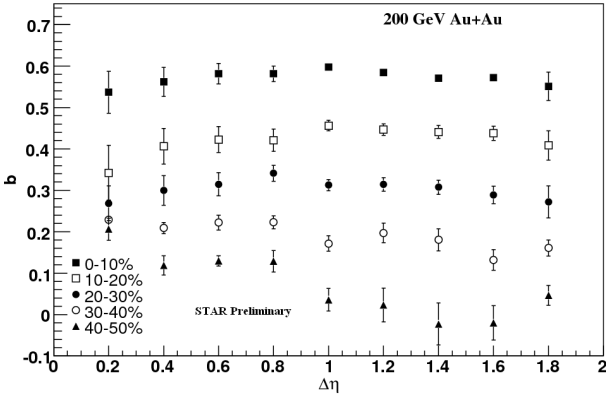}
\caption{The centrality dependence of the FB correlation strength for Au+Au at $\sqrt{s_{NN}}$ = 200 GeV. Most central (0-10\%), 10-20\%, 20-30\%, 30-40\%, and 40-50\%. The long-range correlations persist for the most central Au+Au data, but are similar to the expectation of only short-range correlations beginning at 30-40\% most central data.}
\label{fig2}
\end{figure}

The forward-backward (FB) correlation strength ({\it b}) as a function of centrality from $\sqrt{s_{NN}}$ = 200 GeV Au+Au data is shown in Figure \ref{fig2}. From most central (0-10\%) to mid-peripheral (40-50\%) collisions there is a general decrease in the FB correlation strength. The shape of the FB correlation strength for the 40-50\% bin as a function of $\Delta\eta$ indicates the presence of mostly short-range correlations, which are expected to decrease exponentially with increasing $\eta$ gap. If only short-range correlations contributed to the correlation, all centralities should resemble the 40-50\% result. The presence of a strong correlation that grows with centrality and is flat beyond $\Delta\eta > 1.0$ is indicative of a long-range correlation.

\begin{figure}
\centering
\includegraphics[width=80mm]{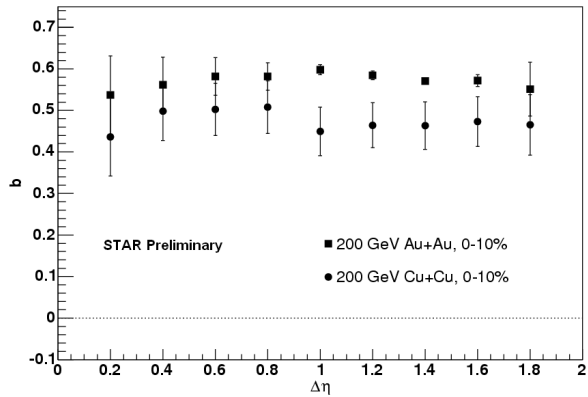}
\caption{Forward-backward correlation strength {\it b} as a function of the pseudorapidity gap $\Delta\eta$ in central (0-10\%) Cu+Cu data (circles) at $\sqrt{s_{NN}}$ = 200 GeV. For comparison, the same quantity is plotted for central (0-10\%) Au+Au data at the same energy (squares).}
\label{fig3}
\end{figure}

The system size dependence of the FB correlation strength is shown in Figure \ref{fig3}. The comparison of the FB correlation strength for most central 0-10\% Au+Au and Cu+Cu collisions at  $\sqrt{s_{NN}}$ = 200 GeV demonstrated that like 0-10\% Au+Au, central Cu+Cu also exhibits a strong, long-range correlation of almost the same magnitude. The plateau of the FB correlation strength in central $\sqrt{s_{NN}}$ = 200 GeV Cu+Cu is about 15\% lower than that of Au+Au at the same energy. However, the multiplicity is $\approx$ 70\% lower, as is the number of participating nucleons (Au+Au, 0-10\% $N_{part} \approx 325$ and Cu+Cu, 0-10\% $N_{part} \approx 100$). The errors bars include statistical and systematic errors.

\begin{figure}
\centering
\includegraphics[width=80mm]{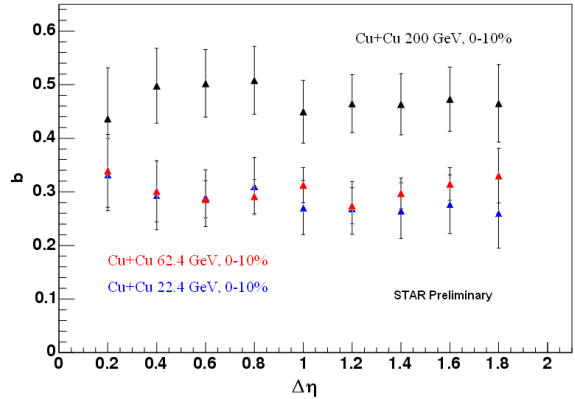}
\caption{The forward-backward correlation strength ({\it b}) as a function of $\Delta\eta$ in the 10\% most central Cu+Cu collisions at energies of $\sqrt{s_{NN}}$ = 200 (black triangles), 62.4 (red triangle), and 22.4 GeV (blue triangles). The correlation strength in Cu+Cu collisions at $\sqrt{s_{NN}}$ = 200 GeV is larger than that at $\sqrt{s_{NN}}$ = 62.4 or 22.4 GeV, both of which are in close agreement.}
\label{fig4}
\end{figure}

Figure \ref{fig4} shows the energy dependence of the FB correlation strength in central 0-10\% Cu+Cu collisions at $\sqrt{s_{NN}}$ = 200, 62.4, and 22.4 GeV. The FB correlation strength is lower in central Cu+Cu collisions at $\sqrt{s_{NN}}$ = 62.4 and 22.4 GeV compared to that at $\sqrt{s_{NN}}$ = 200 GeV, but the lower two energies are in good agreement with each other. Were the FB correlation strength only dependent on the multiplicity of an event, there should be a difference between Cu+Cu at $\sqrt{s_{NN}}$ = 62.4 GeV and 22.4 GeV, as the multiplicities decrease with energy. Further study is ongoing to determine the reasons for such quantitative agreement.

\begin{figure}
\centering
\includegraphics[width=80mm]{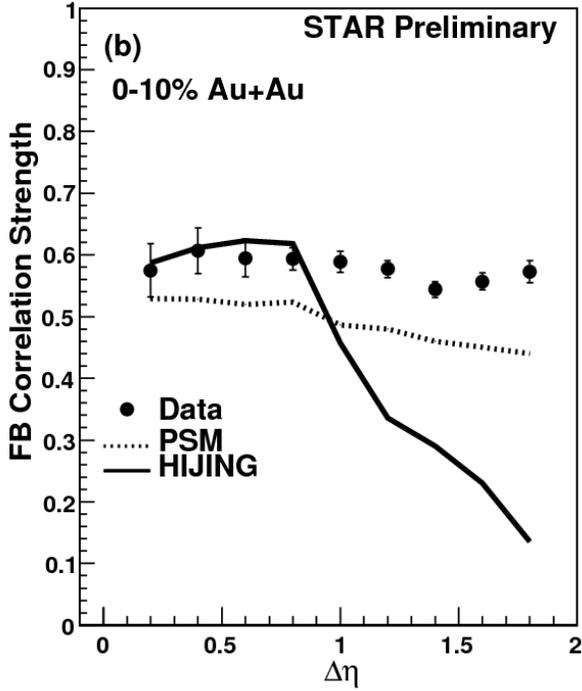}	
\caption{Forward-backward correlation strength from central 0-10\% Au+Au collisions at $\sqrt{s_{NN}}$ = 200 GeV compared with model predictions from the Dual Parton Model (DPM) and HIJING. HIJING is in good agreement with only the short-range component, while the DPM follows the overall trend of the data.}
\label{fig5}
\end{figure}

The $\sqrt{s_{NN}}$ = 200 GeV Au+Au central 0-10\% results for the FB correlation strength were compared with the predictions from the Dual Parton Model (DPM) and HIJING \cite{DPM, HIJING}. The results are shown in Fig. \ref{fig5}. The model output was analyzed following the same procedure as with the data. HIJING predicts a large value of the FB correlation strength near midrapidity, which drops quickly with increasing $\Delta\eta$ due to the absence of long strings in HIJING. In contrast, the FB correlation strength predicted by the DPM has a slower decrease with $\Delta\eta$ due to the inclusion of long strings in the model. The DPM is in qualitative agreement with the data across most of the $\Delta\eta$ range.

\subsection{$K/\pi$ Fluctuations}

\begin{figure}
\centering
\includegraphics[width=80mm]{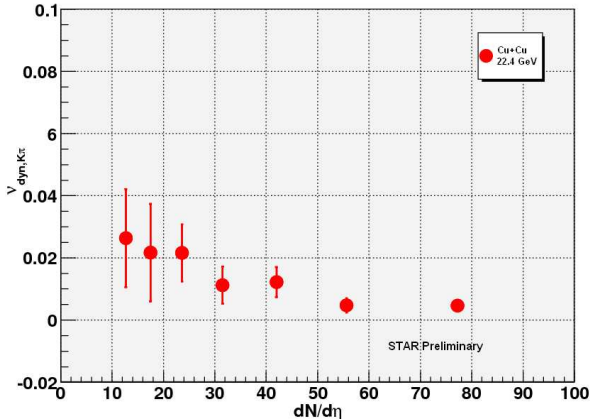}
\caption{$\nu_{dyn,K/\pi}$ as a function of uncorrected charged particle multiplicity $dN/d\eta$ for centralities (with decreasing $dN/d\eta$) 0-5, 5-10, 10-20, 20-30, 30-40, 40-50, and 50-60\% most central Cu+Cu $\sqrt{s_{NN}}$ = 22.4 GeV collisions.}
\label{fig6}
\end{figure}

The centrality dependence of $K/\pi$ fluctuations in $\sqrt{s_{NN}}$ = 22.4 GeV collisions is shown in Fig. \ref{fig6}. $\nu_{dyn,K/\pi}$ as a function of uncorrected $dN/d\eta$ is plotted for the centralities (from right to left) 0-5, 5-10, 10-20, 20-30, 30-40, 40-50, and 50-60\% most central collisions. Previous measurements of $\nu_{dyn,K/\pi}$ as a function of $dN/d\eta$ in Au+Au collisions at $\sqrt{s_{NN}}$ = 62.4 and 200 GeV demonstrate a $1/(dN/d\eta)$ dependence \cite{starkpiprl}. This trend is also seen in Fig. \ref{fig6} for the lower energy and smaller system size in Cu+Cu collisions at $\sqrt{s_{NN}}$ = 22.4 GeV.

\begin{figure}
\centering
\includegraphics[width=80mm]{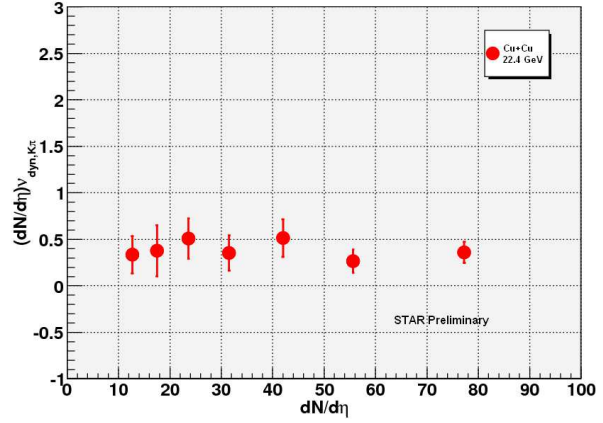}	
\caption{$\nu_{dyn,K/\pi}$ scaled by $dN/d\eta$ as a function of uncorrected charged particle multiplicity $dN/d\eta$ for centralities (with decreasing $dN/d\eta$) 0-5\%, 5-10\%, 10-20\%, 20-30\%, 30-40\%, 40-50\%, and 50-60\% most central Cu+Cu $\sqrt{s_{NN}}$ = 22.4 GeV collisions. $\nu_{dyn,K/\pi}$ is shown to scale linearly with $dN/d\eta$, as was seen from previous measurements of $\nu_{dyn,K/\pi}$ in more peripheral Au+Au collisions \cite{starkpiprl}.}
\label{fig7}
\end{figure}

Figure \ref{fig7} shows $\nu_{dyn,K/\pi}$ scaled by the charged particle multiplicity, $dN/d\eta$, as a function of multiplicity. The scaled $\nu_{dyn,K/\pi}$ in Cu+Cu collisions at $\sqrt{s_{NN}}$ = 22.4 GeV appears to scale with $dN/d\eta$. This behavior was seen for the total charged particle measurement of $\nu_{dyn,K/\pi}$ in more peripheral Au+Au collisions ($dN/d\eta < 400$) at $\sqrt{s_{NN}}$ = 62.4 and 200 GeV \cite{starkpiprl}.

\section{Summary}

Forward-backward multiplicity correlations and particle ratio fluctuations have been discussed. The forward-backward correlation strength ({\it b}) has been shown to be large and approximately flat as a function of $\Delta\eta$ for the most central Au+Au and Cu+Cu collisions at $\sqrt{s_{NN}}$ = 200 GeV. In more peripheral collisions, the forward-backward correlation strength becomes smaller and ultimately demonstrates a behavior consistent with predominantly short-range correlations. The long-range correlations can be ascribed to partonic interactions in the initial state of heavy ion collisions through longitudinal color fields, which are predicted in the Dual Parton Model and the Color Glass Condensate/Glasma description of particle production.

Preliminary results on the fluctuation of the $K/\pi$ ratio, expressed as $\nu_{dyn,K/\pi}$ for Cu+Cu collisions at $\sqrt{s_{NN}}$ = 22.4 GeV have been presented. The general behavior of the $K/\pi$ fluctuations are in agreement with those measured in Au+Au collisions at $\sqrt{s_{NN}}$ = 200 and 62.4 GeV for small $dN/d\eta$. $\nu_{dyn,K/\pi}$ is found to scale with multiplicity as a function of centrality in the lower energy Cu+Cu collisions. Future studies will involve more events and a study of the charge dependence of $\nu_{dyn,K/\pi}$. The dynamic fluctuations of other particle ratios,  $p/\pi$ and $K/p$, will also be studied.


\bigskip 

\end{document}